%% file: UplinkDownlink.tex
\documentclass[10pt,conference]{IEEEtran}
\usepackage{amsmath,amssymb}
\usepackage{graphicx}
\usepackage{mathrsfs}
\usepackage{color}
\usepackage{tikz}
\usepackage{enumerate}
\usepackage{psfrag}	
\usepackage{array}
\usepackage{multirow,hhline}
\usepackage{exscale}
\usepackage{colortbl}
\usepackage{epsfig,subfigure}
\usepackage{cite}
\usepackage{amsthm}

\usepackage[outline]{contour}
\contourlength{1.2pt}
\usepackage{tikz}
\usepackage{pgfplots}
\usepackage{pgfplotstable}
\usepackage{rotate}
\usepgfplotslibrary{units}
\usetikzlibrary{%
patterns,%
calc,%
fit,%
arrows,%
plotmarks,%
shadows,%
chains,%
shapes%
}
\usepackage[outline]{contour}
\contourlength{1.2pt}
\tikzset{>=latex'} 
\tikzstyle{every picture}+=[remember picture] 
\definecolor{rubgray}{cmyk}{0.03,0.03,0.03,0.1}

\newtheorem{theorem}{Theorem}

\newtheorem{proposition}{Proposition}

\newcommand{\x}[0]{\ensuremath{\boldsymbol{x}}}
\newcommand{\y}[0]{\ensuremath{\boldsymbol{y}}}
\newcommand{\z}[0]{\ensuremath{\boldsymbol{z}}}
\renewcommand{\u}[0]{\ensuremath{\boldsymbol{u}}}

\renewcommand{\b}[0]{\ensuremath{\boldsymbol{b}}}

\begin{document}

\IEEEoverridecommandlockouts
\title{Device-Relaying in Cellular D2D Networks: A Fairness Perspective}
\author{
\IEEEauthorblockN{Anas Chaaban and Aydin Sezgin}
\IEEEauthorblockA{Institute of Digital Communication Systems,\\
Ruhr-Universit\"at Bochum (RUB), Germany\\
Email: {anas.chaaban,aydin.sezgin@rub.de}\\
}
\thanks{
The work of A. Chaaban and A. Sezgin is supported by the German Research Foundation, Deutsche Forschungsgemeinschaft (DFG), Germany, under grant SE 1697/5.
}
}

\maketitle

\begin{abstract}
Device-to-Device (D2D) communication is envisioned to be an integral component of 5G networks, and a technique for meeting the demand for high data rates. In a cellular network, D2D allows not only direct communication between users, but also device relaying. In this paper, a simple instance of device-relaying is investigated, and the impact of D2D on fairness among users is studied. Namely, a cellular network consisting of two D2D-enabled users and a base-station (BS) is considered. Thus, the users who want to establish communication with the BS can act as relays for each other's signals. While this problem is traditionally considered in the literature as a multiple-access channel with cooperation in the uplink, and a broadcast channel with cooperation in the downlink, we propose a different treatment of the problem as a multi-way channel. A simple communication scheme is proposed, and is shown to achieve significant gain in terms of fairness (measured by the symmetric rate supported) in comparison to the aforementioned traditional treatment.
\end{abstract}

\section{Introduction}
A cellular network consisting of multiple users who want to communicate with a base-station (BS) is typically studied in the literature as a combination of a multiple-access channel (MAC) \cite{Ahlswede} in the uplink (UL), and a broadcast channel (BC) \cite{Cover} in the downlink (DL). The demand for higher rates of communication has motivated researchers to seek methods to improve the performance of transmission schemes for the MAC and the BC. This research has taken many different directions in the past (code design, MIMO, etc.). Although these directions will remain of great importance for future cellular networks, a new promising direction that will be of high importance in the future is user cooperation.

User cooperation is possible in a cellular network by enabling Device-to-Device communication (D2D). D2D is a technique that is envisioned as a potential solution for meeting the high data-rate demand in the future cellular networks \cite{AsadiWangMancuso}. The main idea of D2D is that devices within close proximity communicate with each other directly, without involving a BS in their communication \cite{LayaWangWidaaZarate}, thus offloading some traffic from the BS. D2D is not only useful for this scenario where the devices act as either a source or a destination, but also for scenarios where devices can act as relays that support other devices' communication with the BS \cite{TehraniUysalYanikomeroglu}. The latter variant, which we term device-relaying cellular network (DRCN), enables user cooperation. It also meets the goal of D2D, which is mainly offloading traffic from the BS. For example, a user with bad channel quality who requires $T$ resource units (time/frequency) to send data directly to the BS, requires $T'<T$ resource units to send the same data through a user with a better channel quality (acting as a relay), which in turn frees $T-T'$ resource units at the BS. Therefore, this enables the BS to support more users; the ultimate goal of designing a network.

User cooperation has been studied earlier in the context of the MAC and the BC \cite{Willems,SendonarisErkipAazhang,KayaUlukus, LiangVeeravalli}. In this paper, we would like to shed light on the following aspect: Interpreting the DRCN as a cooperative MAC or BC is restrictive, in the sense that it restricts communication over a given channel resource (time/frequency) to either UL or DL but not both. Enabling both UL and DL simultaneously over a given channel resource can lead to better performance in terms of achievable rates since the coding scheme can be designed jointly for the UL and DL. Therefore, we propose approaching this problem from another perspective. In particular, we approach the problem from a multi-way channel perspective, where the multi-way channel is an extension of the two-way channel \cite{Shannon_TWC} to multiple users, and has been studied in \cite{ChaabanMaierSezgin,Ong,MaierChaabanMatharSezgin}. 

New opportunities arise when we treat the DRCN as a multi-way channel, namely, multi-way communication and relaying. Multi-way communication refers to simultaneous direct communication between multiple nodes over the same channel resource. In the simplest case with two users, multi-way communication can double the channel capacity as shown in \cite{Han}. Multi-way relaying on the other hand basically refers to the idea of having a relay node compute a function of the received codewords \cite{NazerGastpar} which is simultaneously useful at multiple destinations, and is based on physical-layer network coding \cite{WilsonNarayananPfisterSprintson}. In its simplest form, multi-way relaying in a network with two users can double the capacity of the network in comparison to one-way relaying \cite{NamChungLee_IT,AvestimehrSezginTse}. Using these opportunity can boost the performance of a DRCN in comparison to the traditional cooperative MAC/BC treatment of the problem.

To show the aforementioned gains, we consider a simple DRCN consisting of two D2D-enabled users and a BS. We propose a simple transmission scheme for the network based on a time sharing combination between three components: two-way communication between user 1 and the BS, two-way communication between user 2 and the BS, and two-way relaying through user 1 (we assume that user 1 is the stronger user). While the sum-capacity of this channel has been characterized in \cite{ChaabanMaierSezgin} within a constant gap, the scheme proposed in \cite{ChaabanMaierSezgin} is not fair, since the weaker user is switched off. In practice, it is interesting to maximize the throughput of the network, under a fairness constraint among the users. This corresponds to maximizing the minimum achievable rate, known as max-min fairness. Thus, we focus here on the fair DRCN where rates are allocated to the users equally (symmetric rate). We write the achievable symmetric rate of the proposed scheme which is based in simultaneous UL/DL, and an alternative scheme based on separate UL/DL. Then we compare the achievable rates numerically as a function of the strength of the D2D channel and of the transmit power available. The comparison shows that as the quality of the D2D channel becomes better, the gain achieved by device-relaying increases. Furthermore, the simultaneous UL/DL scheme is simpler than separate UL/DL (which involves block-Markov encoding and backward decoding) and better performing in terms of symmetric rate. This comparison gives guidelines for future cellular network design, regarding the switching on/off of device-relaying capabilities. Namely, if the gain per unit cost (power, complexity, etc.) obtained by device-relaying is higher that a given target value, then the device-relaying functionality is switched on. Otherwise, the DRCN is operated as a combination of a MAC/BC.

The paper is organized as follows. We introduce the notation and system model in Section \ref{Model}. Then, we propose a simultaneous UL/DL scheme based on two-way communication/relaying in Section \ref{Sec:SimULDL}. Next, a separation based scheme which combines cooperative MAC and BC schemes is introduced in Section \ref{Sec:SepULDL}. Finally, the paper is concluded with a numerical evaluation and discussion in Section \ref{Sec:Comp}.

\section{Notation and System Model}
\label{Model}
\subsection{Notation}

Throughout the paper, we use bold-face letters to denote vectors and normal-face letters to denote scalars. The function $C(x)$ is used to denote $\frac{1}{2}\log(1+x)$ for $x\geq0$, and $C^+(x)=\max\{0,C(x)\}$. For $x\in[0,1]$, $\bar{x}$ denotes $1-x$.

\subsection{System Model}
Consider a cellular network consisting of a base station (BS) and two users as shown in Figure \ref{Fig:DRCellular}. Users 1 and 2 want to communicate with the BS (node 3) in both directions, UL and DL. In the UL, user $i\in\{1,2\}$ wants to deliver $\x_{3i}$ to the BS which decodes $\hat{\x}_{3i}$, while in the DL, the BS wants to deliver $\x_{i3}$ to user $i$, which decodes $\hat{\x}_{i3}$. We assume that all nodes have full-duplex capability\footnote{Although the main idea can be extended to the half-duplex case, discussing the full-duplex case is more suitable here since this paper concerns future networks which are expected to have sufficiently good full-duplex capabilities.}, and that the users can establish D2D communication through the channel indicated by $h_3$ in Fig. \ref{Fig:DRCellular}. Note that since the users do not have signals intended to each other, the D2D channel serves relaying purposes only, leading to a DRCN.

Node $i\in\{1,2,3\}$ sends the signal $\x_i$ of length $n$ channel users, with an average power constraint $\|\x_i\|^2\leq n P_i$. The received signals at the three nodes can be written as
\begin{align}
\y_i=h_j\x_k+h_k\x_j+\z_i,
\end{align}
for distinct $i,j,k\in\{1,2,3\}$. Here $\x_i$ is the transmit signal of node $i$, which can in general contain a combination of fresh information and relayed information. That is, $\x_i$ depends on both the information originating at node $i$ ($\x_{ji}$, $j\neq i$), and the past received signals at this node. The variable $h_i\in\mathbb{R}$ denotes the coefficient of the channel between nodes $j$ and $k$. The channel coefficients are assumed to be static, and globally known at all nodes. The noise signal $\z_i\in\mathbb{R}$ is assumed i.i.d. Gaussian with zero mean and unit variance. Note that since all nodes operate in a full-duplex mode (same time and frequency), the channels are reciprocal. We assume without loss of generality that 
\begin{align}
\label{Ordering}
h_2^2\geq h_1^2,
\end{align} 
while the D2D channel $h_3$ is arbitrary.

The goal of this paper is to study the impact of the D2D channel on the achievable rate of communication, subject to a fairness constraint. That is, the rates of all signals appearing in Fig. \ref{Fig:DRCellular} are equal. In the following sections, we consider two variants of communication. In the first variant, the UL and DL phases take place simultaneously, while in the second variant, they are separated. The first variant is interesting by itself due to the potential gain arising from the use of multi-way communication and relaying. The second variant is interesting since it reflects the gain achieved by employing device-relaying in a traditional cellular system with a separated UL/DL. Moreover, the second variant serves as a benchmark for comparing the performance of modern techniques in a DRCN. We start by presenting the transmission scheme for the simultaneous UL/DL variant.

\begin{figure}[t]
\centering
\begin{tikzpicture}[semithick]
\node (u1) at (1.5,2.5) [rectangle, draw, thin, fill=rubgray, minimum width=.8cm, minimum height=.8cm, rotate=0] {1};
\node at ($(u1)+(0,.6)$) {User 1};
\node (u2) at (3,0) [rectangle, draw, thin, fill=rubgray, minimum width=.8cm, minimum height=.81cm, rotate=0] {2};
\node at ($(u2)+(0,-.7)$) {User 2};
\node (bs) at (0,0) [rectangle, draw, thin, fill=rubgray, minimum width=.8cm, minimum height=.8cm, rotate=0] {3};
\node at ($(bs)+(0,-.7)$) {Base-station};

\node (x31) at ($(u1.east)+(1,.2)$) [] {$\x_{31}$};
\draw[->] (x31.west) to  ($(u1.east)+(0,.2)$);
\node (x13h) at ($(u1.east)+(1,-.2)$) [] {$\hat{\x}_{13}$};
\draw[<-] (x13h.west) to  ($(u1.east)+(0,-.2)$);
\node (x32) at ($(u2.east)+(1,.2)$) [] {$\x_{32}$};
\draw[->] (x32.west) to  ($(u2.east)+(0,.2)$);
\node (x23h) at ($(u2.east)+(1,-.2)$) [] {$\hat{\x}_{23}$};
\draw[<-] (x23h.west) to  ($(u2.east)+(0,-.2)$);
\node (xj3) at ($(bs.west)+(-1.3,.2)$) [] {$\x_{13}$, $\x_{23}$};
\draw[->] (xj3.east) to  ($(bs.west)+(0,.2)$);
\node (x3jh) at ($(bs.west)+(-1.3,-.2)$) [] {$\hat{\x}_{31}$, $\hat{\x}_{32}$};
\draw[<-] (x3jh.east) to  ($(bs.west)+(0,-.2)$);

\draw[<->] (bs.east) to node[fill=white] {$h_2$} (u1.south);
\draw[<->] (bs.east) to node[fill=white] {$h_1$} (u2.west);
\draw[<->] (u2.west) to node[fill=white] {$h_3$} (u1.south);
\end{tikzpicture}
\caption{A device-relaying cellular network with two users. Users 1 and 2 want to communicate with the base station in the uplink, which in turn wants to communicate with the users in the downlink. The users can relay information to each other via the D2D channel $h_3$.}
\label{Fig:DRCellular}
\end{figure}
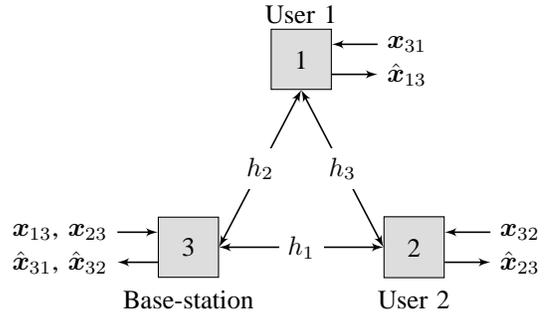

\begin{figure*}
\begin{align}
\label{Rbc}
\bar{R}_{b}&\triangleq \gamma\min\left\{
C^+\left(\frac{h_3^2P_2}{\bar{\alpha}}-\frac{1}{2}\right),
C^+\left(h_2^2P_3-\frac{1}{2}\right),
C\left(\frac{h_2^2P_1}{\bar{\beta}}\right),
C\left(\frac{h_3^2P_1}{\bar{\beta}}\right)\right\}.
\end{align}
\hrule
\end{figure*}

\section{Simultaneous Uplink/Downlink}
\label{Sec:SimULDL}
In a traditional cellular network without device-relaying capabilities, the users would communicate with the BS in a uni-directional manner. That is, users 1 and 2 would send the signals $\x_1=\x_{31}$ and $\x_2=\x_{32}$ e.g. to the BS, respectively, forming a MAC. Similarly, the BS would send the signals $\x_{3}=\x_{13}+\x_{23}$ to users 1 and 2, respectively, forming a BC. The availability of the D2D link $h_3$ between users 1 and 2 enables enhancing this scheme by allowing cooperation between the users \cite{KayaUlukus,LiangVeeravalli}. So far, user cooperation was also applied in a uni-directional manner. That is, a user relays the other user's signal to the BS, or relays the BS signal to the other user. But cooperation can be established in a better way, by making a user relay a signal which is simultaneously useful at the BS and the other user. This can be achieved by physical-layer network coding. Furthermore, if the users and the BS can send information at the same time, then, a channel between a user and the BS can be operated as a two-way channel. With these ideas in mind, several schemes can be designed for the DRCN as described in the following paragraph.

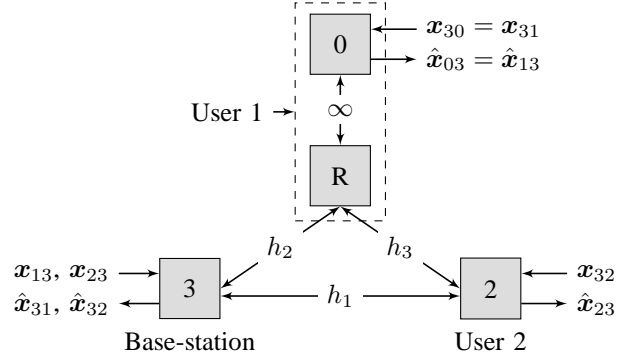
\begin{figure}[t]
\centering
\begin{tikzpicture}[semithick]
\node (rect) at (2,2.35) [rectangle, draw, dashed, thin, minimum width=1.2cm, minimum height=2.9cm, rotate=0] {};
\node (t) at (.5,2.35) {User 1};
\draw[->] (t) to (rect.west);
\node (u1) at (2,3.25) [rectangle, draw, thin, fill=rubgray, minimum width=.8cm, minimum height=.8cm, rotate=0] {0};
\node (r) at (2,1.5) [rectangle, draw, thin, fill=rubgray, minimum width=.8cm, minimum height=.8cm, rotate=0] {R};
\node (u2) at (4,0) [rectangle, draw, thin, fill=rubgray, minimum width=.8cm, minimum height=.81cm, rotate=0] {2};
\node at ($(u2)+(0,-.7)$) {User 2};
\node (bs) at (0,0) [rectangle, draw, thin, fill=rubgray, minimum width=.8cm, minimum height=.8cm, rotate=0] {3};
\node at ($(bs)+(0,-.7)$) {Base-station};

\node (x31) at ($(u1.east)+(1.5,.2)$) [] {$\x_{30}=\x_{31}$};
\draw[->] (x31.west) to  ($(u1.east)+(0,.2)$);
\node (x13h) at ($(u1.east)+(1.5,-.2)$) [] {$\hat{\x}_{03}=\hat{\x}_{13}$};
\draw[<-] (x13h.west) to  ($(u1.east)+(0,-.2)$);
\node (x32) at ($(u2.east)+(1,.2)$) [] {$\x_{32}$};
\draw[->] (x32.west) to  ($(u2.east)+(0,.2)$);
\node (x23h) at ($(u2.east)+(1,-.2)$) [] {$\hat{\x}_{23}$};
\draw[<-] (x23h.west) to  ($(u2.east)+(0,-.2)$);
\node (xj3) at ($(bs.west)+(-1.3,.2)$) [] {$\x_{13}$, $\x_{23}$};
\draw[->] (xj3.east) to  ($(bs.west)+(0,.2)$);
\node (x3jh) at ($(bs.west)+(-1.3,-.2)$) [] {$\hat{\x}_{31}$, $\hat{\x}_{32}$};
\draw[<-] (x3jh.east) to  ($(bs.west)+(0,-.2)$);

\draw[<->] (bs.east) to node[fill=white] {$h_2$} (r.south);
\draw[<->] ($(bs.east)-(0,.1)$) to node[fill=white] {$h_1$} ($(u2.west)-(0,.1)$);
\draw[<->] (u2.west) to node[fill=white] {$h_3$} (r.south);
\draw[<->] (u1.south) to node[fill=white] {$\infty$} (r.north);
\end{tikzpicture}
\caption{The DRCN transformed into a multi-way relay channel with a direct link between two of the users (here user 2 and the BS).}
\label{Fig:EYC}
\end{figure}

A transmission scheme for the DRCN can be obtained from the multi-way relay channel with 3 users \cite{ChaabanSezgin_ISIT12_Y}, where user $i$ wants to send an independent signal $\x_{ji}$ to user $j$ via a relay node by using the following idea. First, the stronger node, say user 1, is split into two nodes, a relay (R) and a virtual user (user 0). Since these two nodes are co-located, we can model them as two separated nodes connected by a channel with infinite capacity. This leads to the representation in Fig. \ref{Fig:EYC}. If $h_1=0$, the problem reduces to a special case of the multi-way relay channel, where users 2 and 0 do not send signals to each other, i.e., $\x_{20}=\x_{02}=0$. Thus, a transmission scheme for this case is readily obtained from the multi-way relay channel. If $h_1\neq 0$, then from a multi-way relay channel point-of-view, we have interference between user 2 and 3 (BS) in Fig. \ref{Fig:EYC}. Even in this case, interference can be dealt with by using backward decoding and interference neutralization as shown in \cite{MaierChaabanMatharSezgin}. Another possibility for communicating over this channel is by switching user 2 off, and operating the DRCN as a two-way channel between user 1 and the BS. Although this scheme achieves the sum-capacity of the channel within a constant gap \cite{ChaabanMaierSezgin}, it is not a fair scheme. In this paper, we propose a fair and simple scheme, which combines schemes for the two-way channel and the two-way relay channel in a TDMA fashion. We present the achievable rate, and then describe its achievability in detail.

\begin{proposition}
\label{Prop:Rates}
The rates $R_{ij}>0$ satisfying 
\begin{align}
R_{13}&\leq \bar{R}_{13},\quad &&R_{23}\leq \bar{R}_b+\bar{R}_{u3},\\
R_{31}&\leq \bar{R}_{31},\quad &&R_{32}\leq \bar{R}_b+\bar{R}_{u2},
\end{align}
are achievable in a DRCN, where 
\begin{align}
\label{R13c}
\bar{R}_{13}&\triangleq \alpha C(h_2^2P_3) ,\\
\label{R31c}
\bar{R}_{31}&\triangleq \alpha C\left(\frac{h_2^2P_1}{\bar{\beta}}\right),\\
\label{Ru23c}
\bar{R}_{u2}&\triangleq \beta C\left(\frac{h_1^2P_2}{\bar{\alpha}}\right),
\end{align}
\begin{align}
\label{Ru32c}
\bar{R}_{u3}&\triangleq \beta C\left(h_1^2P_3\right),
\end{align}
and where $\bar{R}_b$ is defined in \eqref{Rbc} at the top of the page, such that $\alpha+\beta+\gamma= 1$.
\end{proposition}

\begin{proposition}
\label{Prop:ULDLReg}
Simultaneous UL/DL in the DRCN achieves a symmetric rate $R_{31}=R_{32}=R_{13}=R_{23}=R_{\text{sim}}$, where 
$$R_{\text{sim}}=
\max_{\alpha,\beta,\gamma} \min\{\bar{R}_{13},\bar{R}_{31},\bar{R}_b+\bar{R}_{u3},\bar{R}_b+\bar{R}_{u2}\}.$$
\end{proposition}

To achieve $R_{\text{sim}}$, communication over the DRCN is established using the following component schemes. The first component is two-way communication \cite{Han}, which is used for sending signals directly from user 1 to the BS and vice versa, and from user 2 to the BS and vice versa. The second component is two-way relaying through user 1, which is used for sending signals from user 2 to the BS and vice versa, via compute-forward relaying \cite{NamChungLee_IT} at user 1. These components are combined by using TDMA. 

That is, we divide the communication session into three phases: 
\begin{enumerate}
\item Phase 1 spans a fraction $\alpha$ of the total transmission duration, and  is reserved for two-way communication between user 1 and the BS, 
\item phase 2 spans a fraction $\beta$ of the total transmission duration, and is reserved for two-way communication between user 2 and the BS, and 
\item phase 3 spans a fraction $\gamma$ of the total transmission duration, and is reserved for two-way relaying between user 2 and the BS via user 1.
\end{enumerate}

Since user 1 is active in phases 1 and 3, then this user can transmit at a power of $P_1/\bar{\beta}$ without violating the power constraint. Similarly, user 2 can transmit in phases 2 and 3 at a power of $P_2/\bar{\alpha}$. The BS always transmits at a power $P_3$. Next, we describe the transmission procedure in each of the three phases (a summary is given in Table \ref{Tab:Transmit2}).

\begin{table*}
\centering
\begin{tabular}{|c|c|c|c|c|c|c|}
\hline
\multirow{2}{*}{Phase} & \multicolumn{2}{c|}{1} & \multicolumn{2}{c|}{2} & \multicolumn{2}{c|}{3}\\\cline{2-7}
& sends & decodes & sends & decode & sends & decodes\\\hline
User 1 & $\u_{31}$ & $\u_{13}$ & - & - & $\b_1=f(h_2\b_{3}'+h_3\b_{2}')$ & $h_2\b_{3}+h_3\b_{2}$ \\\hline
User 2 & - & - & $\u_{32}$ & $\u_{23}$ & $\b_2$ & $\b_1$ \\\hline
BS & $\u_{13}$ & $\u_{31}$ & $\u_{23}$ & $\u_{32}$ & $\b_3$ & $\b_1$\\\hline
\end{tabular}     
\caption{A summary of the operations at user 1, user 2, and the BS in the three phases of the transmission scheme. The transmitted signal of user 1 in phase 3 is a function $f(\cdot)$ of two-way relaying signals received in previous transmission blocks, denoted $\b_2'$ and $\b_3'$.}
\label{Tab:Transmit2}
\end{table*}

\subsection{Phases 1 and 2}
In phase 1, user 1 and the BS communicate as in a two-way channel \cite{Han}, while user 2 remains silent. That is, user 1 sends $\u_{31}$ and the BS sends $\u_{13}$ with powers $P_1/\bar{\beta}$ and $P_3$, respectively. At the end of the transmission, user 1 decodes $\u_{13}$ and the BS decodes $\u_{31}$. The achievable rate of this phase is as given in the constraints \eqref{R13c} and \eqref{R31c} \cite{Han}, where the factor $\alpha$ accounts for the duration of phase 1. 

Similar transmission is used between user 2 and the BS in phase 2, where user 2 sends $\u_{32}$ and the BS sends $\u_{32}$ with powers $P_2/\bar{\alpha}$ and $P_3$, respectively. This achieves the rates given in \eqref{Ru23c} and \eqref{Ru32c}, where the factor $\beta$ accounts for the duration of phase 2.

\subsection{Phase 3}
In phase 3, all three nodes are active. Namely, user 2 and the BS communicate through user 1 which acts as a bi-directional relay. Thus, user 2 sends a signal $\b_{2}$ with power $P_2/\bar{\alpha}$, and the base station sends a signal $\b_{3}$ with power $P_3$. The two signals have rate $R_b$, and are constructed using a nested-lattice code \cite{NazerGastpar}. User 1 thus is able to decode $h_2\b_{3}+h_3\b_{2}$ \cite{NamChungLee_IT,Nazer_IZS2012}, which is possible if the rate $R_b$ satisfies
\begin{align}
R_b&\leq \min\left\{C^+\left(h_2^2P_3-\frac{1}{2}\right),C^+\left(\frac{h_3^2P_2}{\bar{\alpha}}-\frac{1}{2}\right)\right\}.
\end{align}
Then, user 1 maps the decoded sum to a signal $\b_1$ with power $P_1/\bar{\beta}$ and rate $R_b$, and sends this signal to both user 2 and the BS in the next transmission block. User 2 and the BS can decode this signal if
\begin{align}
R_b&\leq \min\left\{C\left(\frac{h_2^2P_1}{\bar{\beta}}\right),C\left(\frac{h_3^2P_1}{\bar{\beta}}\right)\right\}.
\end{align}
After decoding $\b_1$, user 2 and the BS obtain the sum $h_2\b_{3}+h_3\b_{2}$. User 2 then extracts $\b_{3}$ by subtracting $\b_{2}$, and the BS extracts $\b_{2}$ by subtracting $\b_{3}$. This leads to the rate constraint \eqref{Rbc}, where the factor $\gamma$ accounts for the duration of phase 3.

The combination of these three phases achieves the rates given in Proposition \ref{Prop:Rates}. Since we are seeking a symmetric rate (fair scheme), we calculate the minimum between the achievable rates for a given $\alpha$, $\beta$, $\gamma$, and optimize the outcome over all time sharing parameters satisfying $\alpha+\beta+\gamma=1$. This leads to the rate $R_\text{sim}$ given in Proposition \ref{Prop:ULDLReg}.

In order to evaluate the performance of the given scheme, we compare it with a scheme based on uplink/downlink separation. This scheme is described briefly next.

\section{Uplink/Downlink Separation}
\label{Sec:SepULDL}
The DRCN can be interpreted as a MAC with cooperation (MAC-C) between the transmitters in the UL, and as a BC with cooperation (BC-C) between the receives in the DL. As such, in addition to the simultaneous UL/DL operation mode explained above, the network can be operated as a combination of a MAC-C and a BC-C by separating the UL and DL in time. That is, communication is divided into two phases, a UL MAC-C phase and a DL BC-C phase. These two models have been studied in literature. The following paragraphs review results on the achievable rate regions in the MAC-C and the BC-C.

\subsection{MAC-C}
The capacity of the UL phase has been studied in \cite{SendonarisErkipAazhang, KayaUlukus}, where a transmission scheme based on Willems' results on the MAC with generalized feedback \cite{Willems} was proposed. Shortly, in transmission block $t$, user $i$ sends $\x_i(t)=\x_{ji}(t)+\x_{3i}(t)+\alpha_i\x_{c}(t)$, where $\x_{ji}$ is the cooperation signal form user $i$ to user $j\neq i$ with power $p_{ji}$, $\x_{3i}$ is the signal intended to the BS with power $p_{3i}$, and $\x_{c}$ is a common signal sent by both users to the BS, where the power of $\alpha_i\x_{c}$ is $p_{ci}$. The signal $\x_{c}$ is available at both users as a result of cooperation using $\x_{ji}$. In other words, $\x_{c}(t)$ is a function of the signals $\x_{ji}(t-1)$ and $\x_{ij}(t-1)$ where the latter has been decoded by user $i$ in transmission block $t-1$. The power constraint of user $i$ is satisfied if $p_{ji}+p_{3i}+p_{ci}\leq P_i$. The decoding of cooperation signals is done by decoding $\x_{ji}$ while treating $\x_{3i}$ as noise at user $i$, while using the common known signal $\x_{c}$ as side-information. The BS decodes both the intended signals $\x_{31}$ and $\x_{32}$, and the cooperation signals $\x_{21}$, $\x_{12}$, and $\x_{c}$. The achievable symmetric rate of this scheme is given as follows~\cite{KayaUlukus}.

\begin{theorem}
\label{Thm:MAC-C}
The symmetric uplink rate $R_{31}=R_{32}=R^\mathsf{u}(P_1,P_2)$ is achievable in the MAC-C, where $$R^\mathsf{u}(P_1,P_2)=\max\min\left\{\bar{R}_{1}^\mathsf{u}, \bar{R}_{2}^\mathsf{u}, \frac{1}{2}\bar{R}_{\Sigma,1}^\mathsf{u},\frac{1}{2}\bar{R}_{\Sigma,2}^\mathsf{u}\right\},$$ where the maximization is over all power allocations satisfying $p_{ji}+p_{3i}+p_{ci}\leq P_i$, and where
\begin{align}
\label{RiMACC}
\bar{R}_{i}^\mathsf{u}&\triangleq A_i+C\left(h_j^2p_{3i}\right),\ i,j\in\{1,2\},\ i\neq j\\
\bar{R}_{\Sigma,1}^\mathsf{u}&\triangleq C\left(h_2^2P_1+h_1^2P_2+2\sqrt{h_1^2h_2^2p_{c1}p_{c2}}\right)\\
\bar{R}_{\Sigma,2}^\mathsf{u}&\triangleq C\left(h_2^2p_{31}+h_1^2p_{32}\right)+A_1+A_2,
\end{align}
and $A_i\triangleq C\left(\frac{h_3^2p_{ji}}{1+h_3^2p_{3i}}\right)$.
\end{theorem}

\subsection{BC-C}
The BC-C has been studied in \cite{LiangVeeravalli}, where the capacity of the channel was studied. A transmission scheme for the BC-C was proposed based on a combination of superposition block-Markov encoding, decode-forward at user 1 (the stronger user), and compress-forward at user 2 (the weaker user). Namely, the BS sends $\x_3=\x_{c3}+\x_{23}+\x_{13}$, where $\x_{c3}$ has power $p_{c3}$, $\x_{23}$ has power $p_{23}$, and $\x_{13}$ has power $p_{13}$. The signal $\x_{c3}$ is a cooperation signal, desired at user 2, but also decoded by user 1 for relaying in subsequent transmissions. That is, user 1 decodes $\x_{c3}$ and uses it to generate $\x_1$ with power $P_1$. The signals $\x_{23}$ and $\x_1$ are scaled versions of each other, i.e., $\x_1=\alpha\x_{23}$, which allows increased rates of decoding $\x_{23}$ at user 2. Finally, the signal $\x_{13}$ is dedicated to user 1. User 2 helps user 1 in decoding this latter signal by compressing its received signal $\y_2$, and sending the compressed signal as $\x_2$ with power $p_2$. The power constraints are satisfied if $p_{c3}+p_{23}+p_{13}\leq P_3$, and $p_2\leq P_2$. User 1 decodes $\x_{c3}$ first, then it decodes $\x_2$ and decompresses $\y_2$, and next combines this decompressed $\y_2$ with $\y_1$ for decoding $\x_{13}$. User 2 decodes $\x_{c3}$ and $\x_{23}$. This results in the following achievable symmetric rate \cite{LiangVeeravalli}.
\begin{theorem}
\label{Thm:BC-C}
The symmetric downlink rate $R_{13}=R_{23}=R^\mathsf{d}(P_,P_2,P_3)$ is achievable in a BC-C, where $$R^\mathsf{d}(P_1,P_2,P_3)=\max\min\{\bar{R}_{1}^\mathsf{d},\bar{R}_{2,1}^\mathsf{d},\bar{R}_{2,2}^\mathsf{d}\},$$ where the maximization is over all power allocations satisfying $p_2\leq P_2$ and $p_{c3}+p_{23}+p_{13}\leq P_3$, and where
\begin{align}
\bar{R}_{1}^\mathsf{d}&\triangleq C\left(h_2^2p_{13}+\frac{h_1^2h_3^3p_{13}p_2}{1+h_3^2p_2+(h_1^2+h_2^2)p_{13}}\right)\\
\bar{R}_{2,1}^\mathsf{d}&\triangleq  C\left(\frac{h_1^2(p_{c3}+p_{23})+h_3^2P_1+2\sqrt{h_1^2h_3^2p_{23}P_1}}{1+h_1^2p_{13}}\right)\\
\label{R22BCC}
\bar{R}_{2,2}^\mathsf{d}&\triangleq C\left(\frac{h_2^2p_{c3}}{1+h_2^2p_{13}}\right).
\end{align}
\end{theorem}

\subsection{MAC-BC-C scheme}
From Theorems \ref{Thm:MAC-C} and \ref{Thm:BC-C}, we can design a transmission strategy for the DRCN. Namely, we can state the following achievable symmetric rate.
\begin{proposition}
\label{Prop:MAC-BC-C}
Separate UL/DL in the DRCN achieves a symmetric rate $R_{31}=R_{32}=R_{13}=R_{23}=R_\text{sep}$, where $$R_{\text{sep}}=\max
\min\left\{\tau R^\mathsf{u}(p_1^\mathsf{u},p_2^\mathsf{u}),\bar{\tau} R^\mathsf{d}\left(p_1^\mathsf{d},p_2^\mathsf{d},P_3/\bar{\tau}\right)\right\},$$
where the maximization is over $\tau\in[0,1]$, $p_i^\mathsf{u}\leq P_i$ with $p_i^\mathsf{d}=(P_i-\tau p_i^\mathsf{u})/\bar{\tau}$, $i=1,2$.
\end{proposition}
The transmission scheme achieving the symmetric rate $R_{\text{sep}}$ operates as follows. First, the transmission is divided into two phases, a UL phase whose duration is a fraction $\tau\in[0,1]$ of the overall transmission time, and a DL phase during the remaining transmission time. During the UL phase, users 1 and 2 communicate with the BS using the MAC-C scheme of \cite{KayaUlukus}, with powers $p_1^\mathsf{u}$ and $p_2^\mathsf{u}$. Thus, the uplink symmetric rate given by $R^\mathsf{u}(p_1^\mathsf{u},p_2^\mathsf{u})$ is achievable. During the DL phase, the BS communicates with users 1 and 2 with power $P_3/\bar{\tau}$, which in turn cooperate with powers $p_1^\mathsf{d}$, and $p_2^\mathsf{d}$, respectively. This, the downlink symmetric rate given by $R^\mathsf{d}\left(p_1^\mathsf{d},p_2^\mathsf{d},\frac{P_3}{\bar{\tau}}\right)$ is achievable. The overall symmetric rate is the minimum between $\tau R^\mathsf{u}(p_1^\mathsf{u},p_2^\mathsf{u})$ and $\bar{\tau}R^\mathsf{d}\left(p_1^\mathsf{d},p_2^\mathsf{d},\frac{P_3}{\bar{\tau}}\right)$, maximized over the parameters $\tau$, $p_1^\mathsf{u}$, and $p_2^\mathsf{u}$. It remains to guarantee that the power constraints are satisfied. Since the BS is active for a fraction $\bar{\tau}$ of the time, the power constraint at the BS is satisfied. The power constraints at the users are satisfied by the choice of $p_i^\mathsf{d}$ given $p_i^\mathsf{u}\leq P_i$.

\section{Comparison and Discussion}
\label{Sec:Comp}
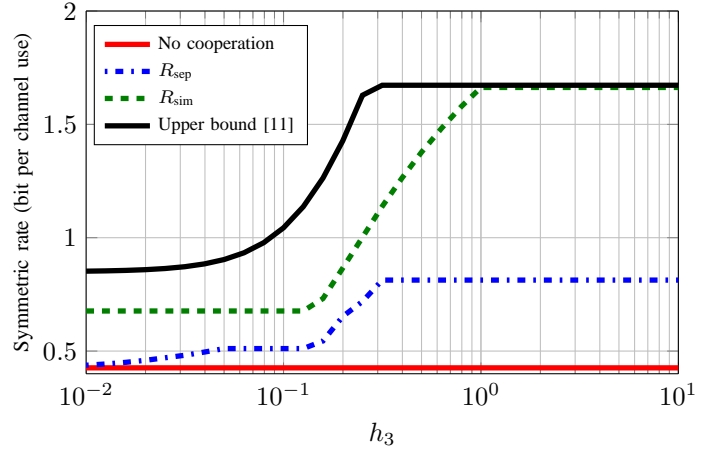
\begin{figure}
\centering
\input{R_vs_h3_P_100_h1_p15_h2_1_tdma}
\caption{Symmetric rates $R_\text{sim}$ and $R_\text{sep}$ as a function of the D2D channel $h_3$, for a DRCN with $P_1=P_2=P_3=100$ and $h_2=1$, $h_1=0.15$. The upper bound is obtained from  \cite{ChaabanMaierSezgin}.}
\label{Fig:Comparison}
\end{figure}

In order to compare the performance of the two schemes, the simultaneous UL/DL and the separate UL/DL, we choose a setup where $h_2=1$, $h_1=0.15$, and $P_1=P_2=P_3=P=100$. That is, the signal-to-noise ratio of the channel $h_2$ is  $\mathsf{SNR}_2=h_2^2P=20$dB and that of the channel $h_1$ is $\mathsf{SNR}_1=h_1^2P=3.5$dB. We calculate the achievable symmetric rates $R_\text{sim}$ and $R_\text{sep}$ numerically, for $h_3\in[0.01,10]$. As we can see in Fig. \ref{Fig:Comparison}, both schemes achieve cooperation gain. This gain increases when $h_3$ becomes larger than the weaker channel $h_1$. Furthermore, the simultaneous UL/DL scheme outperforms the separate UL/DL scheme.

The behaviour of $R_\text{sim}$ can be interpreted as follows. If $h_3$ is small, then relaying through user 1 does not help since the resulting rate \eqref{Rbc} will be close to zero. In this case, $\gamma$ is set to zero, and $\alpha$ and $\beta$ are set so that $\alpha C(h_2^2P)=\beta C(h_1^2P)$ and $\alpha+\beta=1$, leading to a rate given by half the harmonic mean of $C(h_2^2P)$ and $C(h_1^2P)$. As $h_3$ increases beyond $h_1$ but is still smaller than $h_2$, the rate $\bar{R}_b$ becomes larger than $C(h_1^2P)$ and it becomes better for user 2 to communicate with the BS through user 1. In this case, $\beta$ is set to zero, and we need to choose $\alpha$ and $\gamma$ such that $\min\{\bar{R}_b,\alpha C(h_2^2P)\}$ is maximized. Since $\bar{R}_b$ increases with $h_3$, so does $R_\text{sim}$. This increase continues until $h_3$ becomes larger than $h_2$, at which $\bar{R}_b$ stops to increase since $h_2$ becomes the bottleneck. At this point, setting $\alpha=\gamma=\frac{1}{2}$ achieves $R_\text{sim}$ close to optimal. On the other hand, if $h_3$ is small, the MAC-BC-C performs similar to a MAC/BC without cooperation, and its performance improves slowly as $h_3$ increases. Then, the rate $R_\text{sep}$ saturates due to \eqref{R22BCC}. The gain achieved by simultaneous UL/DL is significant especially at large $h_3$ (nearly two-fold), and can be interpreted as a multi-way communication/relaying gain. The plotted upper bound is obtained from \cite{ChaabanMaierSezgin}. 

Simultaneous UL/DL achieves lower gain for lower values of $h_3$ in comparison to MAC/BC without cooperation. This observation is interesting for future cellular network design, since it indicates when the relaying functionality of a device should be switched on. Namely, if the rate-gain is beyond a certain value that is acceptable for a given price (power, complexity, etc.), then the functionality is switched on, otherwise, the network is operated in the traditional MAC/BC mode.

\begin{figure}
\centering
\input{R_vs_P_h1_p5_h2_1_h3_2}
\caption{Symmetric rates $R_\text{sim}$ and $R_\text{sep}$ as a function of $\mathsf{SNR}_2$ for a DRCN with $P_1=P_2=P_3=P$ and $h_1=0.5$, $h_2=1$, and $h_3=2$.}
\label{Fig:FP}
\end{figure}
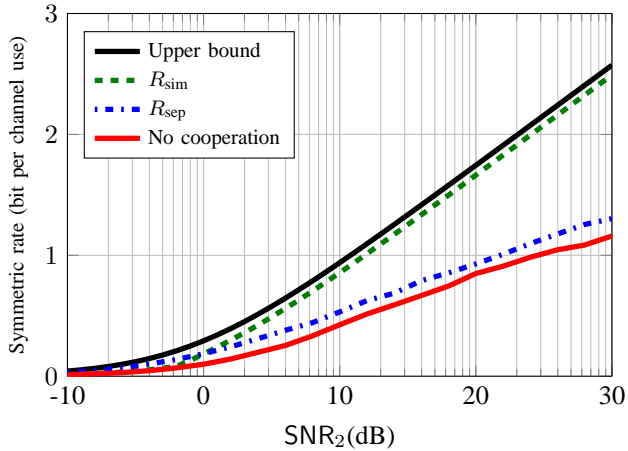

As a conclusion, in a cellular network with D2D communication, where users are allowed to relay information to/from other users, simultaneous UL/DL can achieve higher symmetric rates than separate UL/DL. The achievable symmetric rates are plotted in Fig. \ref{Fig:FP} for a setup with $h_1=0.5$, $h_2=1$, $h_3=2$, and $P=P_1=P_2=P_3$, as a function of $\mathsf{SNR}_2=h_2^2P$. A significant gain is achieved by simultaneous UL/DL in the moderate/high $\mathsf{SNR}$ regime. This multi-way communication/relaying gain can be a potential solution for high data-rate demand in future communication systems, especially that D2D communication is envisioned to be part of those systems.

\bibliography{myBib}

\end{document}

%% file: R_vs_h3_P_100_h1_p15_h2_1_tdma.tex
%
%
\definecolor{mycolor1}{rgb}{0.00000,0.49804,0.00000}%
\begin{tikzpicture}

\begin{axis}[%
width=3.1in,
height=1.9in,
at={(0.758333in,0.489583in)},
scale only axis,
separate axis lines,
every outer x axis line/.append style={black},
every x tick label/.append style={font=\color{black}},
xmode=log,
xmin=0.01,
xmax=10,
xminorticks=true,
xlabel={$h_3$},
xmajorgrids,
xminorgrids,
every outer y axis line/.append style={black},
every y tick label/.append style={font=\color{black}},
ymin=0.4,
ymax=2,
ylabel={\footnotesize Symmetric rate (bit per channel use)},
ylabel style={at={(.05,0.5)}},
ymajorgrids,
legend style={at={(0.013,0.63)},anchor=south west,legend cell align=left,align=left,draw=black},
]
\addplot [color=red,solid,line width=2.0pt]
  table[row sep=crcr]{%
0.01	0.425960149394344\\
0.0125892541179417	0.425960149394344\\
0.0158489319246111	0.425960149394344\\
0.0199526231496888	0.425960149394344\\
0.0251188643150958	0.425960149394344\\
0.0316227766016838	0.425960149394344\\
0.0398107170553497	0.425960149394344\\
0.0501187233627272	0.425960149394344\\
0.0630957344480193	0.425960149394344\\
0.0794328234724281	0.425960149394344\\
0.1	0.425960149394344\\
0.125892541179417	0.425960149394344\\
0.158489319246111	0.425960149394344\\
0.199526231496888	0.425960149394344\\
0.251188643150958	0.425960149394344\\
0.316227766016838	0.425960149394344\\
0.398107170553497	0.425960149394344\\
0.501187233627272	0.425960149394344\\
0.630957344480193	0.425960149394344\\
0.794328234724281	0.425960149394344\\
1	0.425960149394344\\
1.25892541179417	0.425960149394344\\
1.58489319246111	0.425960149394344\\
1.99526231496888	0.425960149394344\\
2.51188643150958	0.425960149394344\\
3.16227766016838	0.425960149394344\\
3.98107170553497	0.425960149394344\\
5.01187233627272	0.425960149394344\\
6.30957344480193	0.425960149394344\\
7.94328234724282	0.425960149394344\\
10	0.425960149394344\\
};
\addlegendentry{\scriptsize No cooperation};

\addplot [color=blue,dash pattern=on 1pt off 3pt on 3pt off 3pt,line width=2.0pt]
  table[row sep=crcr]{%
0.01	0.437526769057454\\
0.0125892541179417	0.443121967042547\\
0.0158489319246111	0.450187701976162\\
0.0199526231496888	0.45910906837699\\
0.0251188643150958	0.470366291636777\\
0.0316227766016838	0.48147631915181\\
0.0398107170553497	0.495701752517763\\
0.0501187233627272	0.510982135301398\\
0.0630957344480193	0.510982135301398\\
0.0794328234724281	0.510982135301398\\
0.1	0.510982135301398\\
0.125892541179417	0.510982135301398\\
0.158489319246111	0.544579980485838\\
0.199526231496888	0.652610024625653\\
0.251188643150958	0.718881698402157\\
0.316227766016838	0.813084901146679\\
0.398107170553497	0.813084901146679\\
0.501187233627272	0.813084901146679\\
0.630957344480193	0.813084901146679\\
0.794328234724281	0.813084901146679\\
1	0.813084901146679\\
1.25892541179417	0.813084901146679\\
1.58489319246111	0.813084901146679\\
1.99526231496888	0.813084901146679\\
2.51188643150958	0.813084901146679\\
3.16227766016838	0.813084901146679\\
3.98107170553497	0.813084901146679\\
5.01187233627272	0.813084901146679\\
6.30957344480193	0.813084901146679\\
7.94328234724282	0.813084901146679\\
10	0.813084901146679\\
};
\addlegendentry{\scriptsize $R_\text{sep}$};

\addplot [color=mycolor1,dashed,line width=2.0pt]
  table[row sep=crcr]{%
0.01	0.676775007820155\\
0.0125892541179417	0.676775007820155\\
0.0158489319246111	0.676775007820155\\
0.0199526231496888	0.676775007820155\\
0.0251188643150958	0.676775007820155\\
0.0316227766016838	0.676775007820155\\
0.0398107170553497	0.676775007820155\\
0.0501187233627272	0.676775007820155\\
0.0630957344480193	0.676775007820155\\
0.0794328234724281	0.676775007820155\\
0.1	0.676775007820155\\
0.125892541179417	0.676775007820155\\
0.158489319246111	0.731715857185719\\
0.199526231496888	0.865833600961574\\
0.251188643150958	1.00206082815415\\
0.316227766016838	1.13815300253167\\
0.398107170553497	1.26312142282068\\
0.501187233627272	1.37824977692962\\
0.630957344480193	1.48219269472686\\
0.794328234724281	1.57799612141218\\
1	1.66276292279473\\
1.25892541179417	1.66276292279473\\
1.58489319246111	1.66276292279473\\
1.99526231496888	1.66276292279473\\
2.51188643150958	1.66276292279473\\
3.16227766016838	1.66276292279473\\
3.98107170553497	1.66276292279473\\
5.01187233627272	1.66276292279473\\
6.30957344480193	1.66276292279473\\
7.94328234724282	1.66276292279473\\
10	1.66276292279473\\
};
\addlegendentry{\scriptsize $R_\text{sim}$};

\addplot [color=black,solid,line width=2.0pt]
  table[row sep=crcr]{%
0.01	0.852435982228176\\
0.0125892541179417	0.853729028915227\\
0.0158489319246111	0.855773633759552\\
0.0199526231496888	0.859002289636591\\
0.0251188643150958	0.864089945901722\\
0.0316227766016838	0.872080547785205\\
0.0398107170553497	0.884566231129213\\
0.0501187233627272	0.903922397950681\\
0.0630957344480193	0.933574056110289\\
0.0794328234724281	0.978206288289962\\
0.1	1.04373142062517\\
0.125892541179417	1.13674200975192\\
0.158489319246111	1.26327061241545\\
0.199526231496888	1.42710474127002\\
0.251188643150958	1.62847312276419\\
0.316227766016838	1.67249949285486\\
0.398107170553497	1.67249949285486\\
0.501187233627272	1.67249949285486\\
0.630957344480193	1.67249949285486\\
0.794328234724281	1.67249949285486\\
1	1.67249949285486\\
1.25892541179417	1.67249949285486\\
1.58489319246111	1.67249949285486\\
1.99526231496888	1.67249949285486\\
2.51188643150958	1.67249949285486\\
3.16227766016838	1.67249949285486\\
3.98107170553497	1.67249949285486\\
5.01187233627272	1.67249949285486\\
6.30957344480193	1.67249949285486\\
7.94328234724282	1.67249949285486\\
10	1.67249949285486\\
};
\addlegendentry{\scriptsize Upper bound \cite{ChaabanMaierSezgin}};

\end{axis}
\end{tikzpicture}%

%% file: R_vs_P_h1_p5_h2_1_h3_2.tex
%
%
\definecolor{mycolor1}{rgb}{0.00000,0.49804,0.00000}%
\begin{tikzpicture}

\begin{axis}[%
width=2.85in,
height=1.9in,
at={(1.018333in,0.517491in)},
scale only axis,
separate axis lines,
every outer x axis line/.append style={black},
every x tick label/.append style={font=\color{black}},
xmode=log,
xmin=0.1,
xmax=1000,
xtick={0.1,1,10,100,1000},
xticklabels={{-10},{0},{10},{20},{30}},
xminorticks=true,
xlabel={$\mathsf{SNR}_2$(dB)},
xmajorgrids,
xminorgrids,
every outer y axis line/.append style={black},
every y tick label/.append style={font=\color{black}},
ymin=0,
ymax=3,
ylabel={\footnotesize Symmetric rate (bit per channel use)},
ylabel style={at={(.08,0.5)}},
ymajorgrids,
legend style={at={(0.03,0.6)},anchor=south west,legend cell align=left,align=left,draw=black},
]
\addplot [color=black,solid,line width=2.0pt]
  table[row sep=crcr]{%
0.1	0.0424812503605781\\
0.125892541179417	0.0527111911079887\\
0.158489319246111	0.0651905888923648\\
0.199526231496888	0.080311107363757\\
0.251188643150958	0.0984874222830448\\
0.316227766016838	0.120139883420344\\
0.398107170553497	0.145671264424369\\
0.501187233627272	0.175439166924116\\
0.630957344480193	0.209727190545796\\
0.794328234724281	0.248719192951674\\
1	0.292481250360578\\
1.25892541179417	0.340954915407706\\
1.58489319246111	0.393963187852168\\
1.99526231496888	0.451227942677442\\
2.51188643150958	0.512395324214732\\
3.16227766016838	0.577064519202225\\
3.98107170553497	0.644815534495049\\
5.01187233627272	0.715232776983065\\
6.30957344480193	0.787922774880995\\
7.94328234724282	0.862525770916087\\
10	0.938721875540867\\
12.5892541179417	1.01623294593039\\
15.8489319246111	1.09482145428294\\
19.9526231496888	1.17428747391835\\
25.1188643150958	1.25446467373085\\
31.6227766016838	1.33521595863651\\
39.8107170553497	1.41642917266028\\
50.1187233627272	1.4980131095328\\
63.0957344480193	1.57989395317847\\
79.4328234724281	1.66201218944788\\
100	1.74431998087498\\
125.892541179417	1.82677896881989\\
158.489319246111	1.90935845459937\\
199.526231496888	1.9920339075723\\
251.188643150958	2.07478574979622\\
316.227766016838	2.15759837127291\\
398.107170553497	2.24045933536912\\
501.187233627272	2.3233587397908\\
630.957344480193	2.40628870398675\\
794.328234724281	2.48924295880619\\
1000	2.57221651854146\\
};
\addlegendentry{\footnotesize Upper bound};

\addplot [color=mycolor1,dashed,line width=2.0pt]
  table[row sep=crcr]{%
0.1	0.0141426921630963\\
0.125892541179417	0.0177056558362402\\
0.158489319246111	0.0221715427600718\\
0.199526231496888	0.0276900266033626\\
0.251188643150958	0.0345314475899196\\
0.316227766016838	0.0429753454860147\\
0.398107170553497	0.0533158347811064\\
0.501187233627272	0.0659366917251016\\
0.630957344480193	0.0812488751420586\\
0.794328234724281	0.129046905256159\\
1	0.1845\\
1.25892541179417	0.240417691794594\\
1.58489319246111	0.298682818005715\\
1.99526231496888	0.35947968024854\\
2.51188643150958	0.423159485668696\\
3.16227766016838	0.489721817266047\\
3.98107170553497	0.559353625530274\\
5.01187233627272	0.630410873169438\\
6.30957344480193	0.704345147410846\\
7.94328234724282	0.779138290494561\\
10	0.856209325612731\\
12.5892541179417	0.93356980302663\\
15.8489319246111	1.012534430874\\
19.9526231496888	1.0928756828734\\
25.1188643150958	1.17212322162782\\
31.6227766016838	1.25388393624757\\
39.8107170553497	1.33504360048503\\
50.1187233627272	1.41610708550032\\
63.0957344480193	1.49771452446613\\
79.4328234724281	1.58017903635881\\
100	1.66276292279473\\
125.892541179417	1.74544187945975\\
158.489319246111	1.82819650990876\\
199.526231496888	1.91101134940484\\
251.188643150958	1.99387407739335\\
316.227766016838	2.07677488422435\\
398.107170553497	2.15970596321479\\
501.187233627272	2.24266110406562\\
630.957344480193	2.32563536792814\\
794.328234724281	2.40862482805069\\
1000	2.49162636297644\\
};
\addlegendentry{\footnotesize $R_\text{sim}$};

\addplot [color=blue,dash pattern=on 1pt off 3pt on 3pt off 3pt,line width=2.0pt]
  table[row sep=crcr]{%
0.1	0.031093659727998\\
0.158489319246111	0.046291095767607\\
0.251188643150958	0.069360961874892\\
0.398107170553497	0.100339665189072\\
0.630957344480193	0.138158957735719\\
1	0.186472119473808\\
1.58489319246111	0.243460395528974\\
2.51188643150958	0.310599292669899\\
3.98107170553497	0.379462992927842\\
6.30957344480193	0.442877808434556\\
10	0.531838239123287\\
15.8489319246111	0.6255\\
25.1188643150958	0.686684899040295\\
39.8107170553497	0.79\\
63.0957344480193	0.8547\\
100	0.92865939316468\\
158.489319246111	1.00824841686991\\
251.188643150958	1.0966\\
398.107170553497	1.1763\\
630.957344480193	1.2552\\
1000	1.3029\\
};
\addlegendentry{\footnotesize $R_\text{sep}$};

\addplot [color=red,solid,line width=2.0pt]
  table[row sep=crcr]{%
0.1	0.01280669802487\\
0.158489319246111	0.019555998023463\\
0.251188643150958	0.029649950302209\\
0.398107170553497	0.045642194947436\\
0.630957344480193	0.068730720691446\\
1	0.098861608480291\\
1.58489319246111	0.142622779398266\\
2.51188643150958	0.197961090369751\\
3.98107170553497	0.254007650931189\\
6.30957344480193	0.333857440150577\\
10	0.425435126420902\\
15.8489319246111	0.514592165876774\\
25.1188643150958	0.589688498638686\\
39.8107170553497	0.66742898161923\\
63.0957344480193	0.745949739386299\\
100	0.849146890331373\\
158.489319246111	0.91\\
251.188643150958	0.983854860630457\\
398.107170553497	1.0457\\
630.957344480193	1.0852\\
1000	1.1601\\
};
\addlegendentry{\footnotesize No cooperation};

\end{axis}
\end{tikzpicture}%